\documentclass[orivec,11pt]{llncs}
    \usepackage[margin=1.5in]{geometry}
    \usepackage{amsmath,stmaryrd,amssymb}

        \usepackage{amsthm}
    \usepackage{url,xspace,enumerate,epic,eepic,verbatim,hyperref,color}
    \usepackage[dvipsnames]{xcolor}
    \usepackage[utf8]{inputenc}
    \usepackage[final,stretch=10,shrink=20]{microtype}
    \usepackage{mathptmx}
    \usepackage{bm}
    \usepackage[linesnumbered,ruled,vlined]{algorithm2e}
    \usepackage{calrsfs}
\usepackage{todonotes}
\DeclareMathAlphabet{\pazocal}{OMS}{zplm}{m}{n}

\newcommand{\Lv}{\pazocal{V}}
\newcommand{\Lp}{\pazocal{P}}

\newcommand{\DW}{Wikstr\"om}

\newcommand{\?}{\stackrel{?}{=}}

\newtheorem{prop}{Proposition}

    \def\ZZ{\mathbb Z}

    \title{A Description and Proof of a Generalised and Optimised Variant of \DW's Mixnet}
    \author{Thomas Haines}
    \institute{Polyas GmbH}
\begin{document} 
\maketitle
\section{Introduction}
In this paper, we describe an optimised variant of \DW's mixnet which shuffles vectors of ElGamal ciphertexts in parallel. We then show in detail that this construction is secure.\par
A verifiable shuffle takes a packet of ciphertexts, which it re-encrypts and shuffles to produce
an output packet. 
 More specifically, a cryptographic shuffle of ElGamal encryptions $\textbf{e} = (e_1,....,e_N)$ is another list of ElGamal encryptions $\textbf{e}' = (e'_1,...,e'_N)$, which contains the same plaintexts $m_i$ in permuted order.
Given $\textbf{e}$ and $\textbf{e}'$ we may wish to prove that they have this relationship, this called a proof of shuffle.   

\DW's verifiable mixnet as we refer to it here was first presented in ``Proofs of Restricted Shuffles'' by Terelius and \DW  \cite{TereliusWikstroem-AFRICACRYPT-2010}, building on \DW's previous work in \cite{wikstrom2009commitment}.
Specifically we take the optimised variant for ElGamal which appears to be in common use; for instance, it is presented in Haenni et al's pseudo-code algorithms for implementing \DW's verifiable mixnet \cite{haenni2017pseudo}. 
 We extend the mixnet to support parallel shuffles, where each $e_i$ and $e'_i$ are themselves vectors of related ciphertexts. 
 (The possibility of doing this is proven by the \DW's result but we wish to show that this particular instance with its optimisations is secure.)
\section{Notation}

\begin{itemize}
\item 
    $G_q$ is a cyclic group of prime order $q$ in which both the
    decisional and computational Diffie-Hellman problems are
    hard. We will use the multiplicative notation for the group
    operation. As usually, by $\mathbb{Z}_q$ we denote the field of
    integers modulo $q$.
\item 
    $A^N$ is the set of vectors of length $N$
    containing elements of $A$.  We will denote vectors in
    bold, for instance \textbf{a}. We will denote the $i$th
    element using subscript; for instance as $\textbf{a}_i$.
\item 
    Similarly, $A^{N\times N}$ is the set of square matrices of
    order $N$ containing elements of $A$. We will denote matrices
    using upper case letters, for instance $M$.  We will denote the $i$th column
    of $M$ as $M_i$ and the element $i$th row and $j$th column as $M_{i,j}$.  

    A matrix $M$, containing only $0$ and $1$ values, is a
    \emph{permutation matrix}, if every column and every row
    contains exactly one $1$.
\item 
    $PC_{h,h_1}(m,r)$, for $m,r \in \mathbb{Z}_q$ and $h, h_1 \in
    G_q$, is defined as
    $h^r h_1^m$ (note that $h$ and $h_1$ are group elements and
    hence the multiplication here denotes the group
    multiplication). $PC_{h,h_1}(m,r)$ is known as a Pedersen
    commitment.
\item 
    $EPC_{h,h_1,...,h_N}(\textbf{m},r)$, for $\textbf{m} \in
    \mathbb{Z}_q^N$ and $r \in \mathbb{Z}_q$, is defined as 
    $h^r\prod_{i = 1}^N h_i^{\textbf{m}_i}$ 
    (otherwise known as an extended Pedersen commitment).

\item 
    $C_{h,h_1,...,h_N}(M,\textbf{r})$, for $M \in
    \mathbb{Z}_q^{N\times N}$ and $\textbf{r} \in
    \mathbb{Z}_q^N$,
    is $(\textbf{c}_1,...,\textbf{c}_n)$ where 
    $\textbf{c}_i = h^{\textbf{r}_i}\prod_{j=1}^N
    h_j^{M_{j,i}}$, which means that $\textbf{c}_i$ is the
    extended Pedersen commitment to the $i$th column of $M$.

\item 
    $\textbf{Enc}_{g,pk}(m,r)$ for $m \in G_q$ and $r \in
    \mathbb{Z}_q$ is $(g^r, pk^rm)$ (the ElGamal encryption
    of the group element $m$)

\item  
    $\textbf{ReEnc}_{g,pk}(e,r)$, for $e \in G^{2}_q$ and $r
    \in \mathbb{Z}_q$ is $(e_1g^{r}, e_2 \mit{pk}^{r})$.

\item 
    $\textbf{Enc}_{g,pk}(\textbf{m},\textbf{r})$, for $\textbf{m}
    \in G_q^w$ and $\textbf{r} \in \mathbb{Z}_q^w$,
    is $\textbf{Enc}_{g,pk}(\textbf{m}_1, \textbf{r}_1), \dots,
        \textbf{Enc}_{g,pk}(\textbf{m}_w, \textbf{r}_w)$
\item  
    $\textbf{ReEnc}_{g,pk}(\textbf{e},\textbf{r})$, for
    $\textbf{e} \in (G^{2}_q)^w$ and $\textbf{r}  \in
    \mathbb{Z}_q^w$,
    is $\textbf{ReEnc}_{g,pk}(\textbf{e}_1, \textbf{r}_1), \dots,
        \textbf{ReEnc}_{g,pk}(\textbf{e}_w, \textbf{r}_w)$
\item 
    $\langle \textbf{a}, \textbf{b} \rangle$, for $\textbf{a} \in
    \mathbb{Z}_q^N$ and $\textbf{b} \in \mathbb{Z}_q^N$ is $\sum_{i=1}^N
    \textbf{a}_i\textbf{b}_i \mod{q}$.

\item
	$AB$, for $A \in \ZZ_q^{n \times m}$ and $B \in \ZZ_q^{m \times w}$,
	 is a matrix in $\ZZ_q^{n \times w}$ where the value in the $i$th row
	and $j$th column is equal $\sum_{k=1}^m A_{i,k}B_{k,j}$
\item 
	$M\textbf{x}$, for $M \in \ZZ_q^{M \times N}$ and $\textbf{x} \in \ZZ_q^N$,
	 is a vector of length $M$ where $i$th position is equal to $\sum_{j=1}^N M_{i,j}	
	 \textbf{x}_j$. (Note that both this and the next definition are consistent with treating 
	$\textbf{x}$ as a column, and row vector respectively, and applying the definition of matrix 
	multiplication definition from above.

\item 
	$\textbf{x}M$, for $M \in \ZZ_q^{N \times M}$ and $\textbf{x} \in \ZZ_q^N$, is a vector
	of length $M$ where the $i$th position is equal to $\sum_{j=1}^N \textbf{x}_jM_{j,i}$.
	
\item 
	$a\textbf{x}$, for $a \in \ZZ_q$ and $\textbf{x} \in \ZZ_q^N$, is a vector of length
	 $N$ where $i$th position is equal to $a\textbf{x}_i$.
	 
\item 
	$\textbf{x}^a$, for $a \in \ZZ_q$ and $\textbf{x} \in \ZZ_q^N$, is a vector of length
	 $N$ where $i$th position is equal to $\textbf{x}_i^a$.
	
\item
	For two vectors $\textbf{x}, \textbf{y} \in \ZZ_q^N$ we some sometimes abuse notation
	by writing $\textbf{x}+\textbf{y}$, $\textbf{x}*\textbf{y}$, and $\textbf{x}^\textbf{y}$ to denote the
	pairwise addition, multiplication, and exponentiation of the vectors respectively.

\item 
    For a matrix $M$, by $\pi_M$ we denote the permutation of the
    set $\{1, \dots, N \}$ defined by $M$, that is such a
    permutation that for each vector $\textbf{x}$ we have
    $\textbf{x} = (\textbf{y}_{\pi(1)}, \dots, \textbf{y}_{\pi(N)})$,
    where $\textbf{y} = M \textbf{x}$.

\item
    A binary relation $\mathcal{R}$ for a set statements of $S$ and witnesses $W$ is a subset of the cartesian product of $S$ and $W$.

\item
    For two binary relations $\mathcal{R}$ and $\mathcal{R}'$, 
	we denote by $\mathcal{R} \land \mathcal{R}'$ a relation between
    $(S*S')$ and $(W*W')$ the cartesian product	of the statements and 
	witness of $\mathcal{R}$ and $\mathcal{R'}$.  
	The relation is said to hold when both the subrelations hold.

\item
   For two binary relations $\mathcal{R}$ and $\mathcal{R}'$,
	we denote by $\mathcal{R} \lor \mathcal{R}'$ a relation between $(S*S')$ and $(W*W')$.  
	The relation is said to hold when either subrelations holds.

\item
    For two binary relations $\mathcal{R}$ and $\mathcal{R}'$ where $W = W'$, we denote by 
	$\mathcal{R} \underline{\land} \mathcal{R}'$ a relation between $(S*S')$ and $(W)$.
	The relation is said to hold when both the subrelations hold.  We will abuse notation by
	writing $\mathcal{R} \underline{\land} \mathcal{R}'$ when $W \neq W'$ but are both
	cartesian products with subgroups in common.

\end{itemize}

\section{Shuffle Proof - Description and Proof}

\begin{algorithm}
    \SetKwInOut{CommonInput}{Common Input}
    \SetKwInOut{PrivateInput}{Private Input}
    \DontPrintSemicolon

    \CommonInput{%
        A group generator $g \in G_q$, public key $pk
        \in G_q$, matrix commitment $\textbf{c} \in G^N_q$,
        commitment parameters $h, h_1,...,h_N \in G_q$, ciphertext
        vectors $\textbf{e}_1,...,\textbf{e}_N \in {(G^{2}_q)}^w$ and
        $\textbf{e}'_1,...,\textbf{e}'_N \in {(G^{2}_q)}^w$.
    }

    \PrivateInput{%
        Permutation matrix $M \in \mathbb{Z}_q^{N\times
    N}$, randomness $\textbf{r} \in \mathbb{Z}_q^N$ and
    randomness
    $R \in \ZZ_q^{w\times N}$,
    such that $\textbf{c} = C_{h,h_1,...,h_N}(M,\textbf{r})$ and $\textbf{e}'_{i} =
    \textbf{ReEnc}_{g,pk}(\textbf{e}_{\pi_M(i)},R_{\pi_M(i)})$.}

    \BlankLine
    $\Lv$ chooses $\textbf{u} \in \mathbb{Z}^N_q$ randomly and
    hands $\textbf{u}$ to $\Lp$.\;

    $\Lp$ computes $\textbf{u}' = M\textbf{u}$. Then $\Lp$ chooses
    $\hat{\textbf{r}}  \in \mathbb{Z}^N_q$ at random and computes
    \begin{align*}
        \bar{r} &= \textbf{r}_1 +\dots+ \textbf{r}_N,
        &\tilde{r} &= \langle \textbf{r}, \textbf{u}\rangle, \\
        r^\diamond &= \hat{\textbf{r}}_N +
        \sum^{N-1}_{i=1}\left(\hat{\textbf{r}}_i\prod^{N}_{j=i+1}\textbf{u}'_{j}\right),
        &\textbf{r}^\star &= R\textbf{u}
    \end{align*}
    $\Lp$ randomly chooses $\boldsymbol{\hat{\omega}},\boldsymbol{\omega'} \in \mathbb{Z}^N_q$, 
    $\omega_1, \omega_2, \omega_3 \in \mathbb{Z}_q$, and
    $\boldsymbol{\omega_4} \in \mathbb{Z}_q^w$, 
    and hands the following values to $\Lv$:
    \[
    \begin{array}{cccc}
        \hat{c}_0 = h_1, \ \ 
        \hat{c}_i = 
        h^{\hat{\textbf{r}}_i}\hat{c}_{i-1}^{\textbf{u}'_i} \ \ \ (i \in \{1, \dots, N\})
        &\quad t_1 = h^{\omega_1} 
        &\quad t_2 = h^{\omega_2} 
        &\quad t_3 = h^{\omega_3}\prod^N_{i=1}h_i^{\boldsymbol{\omega'}_i} 
        \end{array}
    \]
    \[
    \begin{array}{cccc}
        \textbf{t}_{4} = \textbf{ReEnc}_{g,pk}(\prod^N_{i=1} 
             \textbf{e}_{i}'^{\boldsymbol{\omega'}_i}, -\boldsymbol{\omega}_{4}) 
        &\quad 
             \hat{\textbf{t}}_i = 
               h^{\boldsymbol{\hat{\omega}_i}}\hat{c}_{i-1}^{\boldsymbol{\omega}'_i} \ \ \ (i \in \{1, \dots, N\})
    \end{array}
    \]
      \;
      $\Lv$ chooses a challenge $c \in \mathbb{Z}_q$ at random and sends it to $\Lp$.\;
      $\Lp$ then responds,
      \[
    \begin{array}{cccc}
        s_1 = \omega_1 + c \cdot \bar{r} &\hspace{0.5cm} s_2 = \omega_2 + c \cdot r^\diamond &\hspace{0.5cm}  s_3 = \omega_3 + c \cdot \tilde{r} &\hspace{0.5cm} \textbf{s}_{4} = \boldsymbol{\omega_{4}} + c \cdot \textbf{r}^\star \\
        \end{array}
    \]
    \[
        \begin{array}{cccc}
        \hat{\textbf{s}} = \boldsymbol{\hat{\omega}} + c \cdot \hat{\textbf{r}} &\hspace{0.5cm} \textbf{s}' = \boldsymbol{\omega'} + c \cdot \textbf{u}'
    \end{array}
    \]
      \;
      $\Lv$ accepts if and only if
        \[
    \begin{array}{ccc}
        t_1 = (\prod^N_{i=1} \textbf{c}_i / \prod^N_{i=1}h_i)^{-c}h^{s_1}&\hspace{0.5cm} t_2 = (\hat{c}_N/h_1^{\prod^N_{i=1}\textbf{u}_i})^{-c}h^{s_2} &\hspace{0.5cm}  t_3 = (\prod^N_{i=1}\textbf{c}_i^{\textbf{u}_i})^{-c}h^{s_3}\prod^N_{i=1}h_i^{\textbf{s}'_i}   \\
        \end{array}
    \]
    \[ 
      \begin{array}{cc}
         \textbf{t}_{4} = \textbf{ReEnc}_{g,pk}((\prod^N_{i=1}\textbf{e}_{i}^{\textbf{u}_i})^{-c}\prod^N_{i=1}\textbf{e}'^{\textbf{s}'_i}_{i},-\textbf{s}_{4}) &\hspace{0.5cm} \hat{\textbf{t}}_i = \hat{c}_i^{-c}h^{\hat{\textbf{s}}_i}\hat{c}_{i-1}^{\textbf{s}'_i}\\
    \end{array}
      \]
      \;

  \caption{Interactive ZK-Proof of Extended Shuffle}\label{algo:shuffle}
\end{algorithm}

\newpage

\subsubsection{Formal Security Statement}

In the security statement for the presented shuffle algorithm, we
will use the following notation.

\begin{itemize}
\item 
    $\mathcal{R}_{com}(h,h_1,...,h_N)(\textbf{m},r,\textbf{m}',r')$ is a relationship
    between the commitment parameters $(h,h_1,...,h_N)$ and $(\textbf{m},\textbf{m}' \in \ZZ_q^{N} r,r' \in \ZZ_q)$  which holds if and only if EPC($\textbf{m},r$) = EPC($\textbf{m}',r'$) and $\textbf{m} \neq \textbf{m}'$.
\item 
    $\mathcal{R}_{\pi}(h,h_1,...,h_N,c)(M,\textbf{r})$ is the
    relationship between the commitment parameters
    $(h,h_1,...,h_N)$, a commitment $c \in G_q$, a permutation matrix $M \in \ZZ_q^{N\times N}$,
    and $\textbf{r} \in \ZZ_q^N$ which holds if
    $C_{h,h_1,...,h_N}(M,\textbf{r})=c$.

\item 
    $\mathcal{R}^{shuf}_{ReEnc_{g,pk}}(g,pk,
    (\textbf{e}_1,...,\textbf{e}_N),
    (\textbf{e}'_1,...,\textbf{e}'_N))(\pi_M,
    (\textbf{r}'_1,...,\textbf{r}'_N))$, where $\pi_M$ is a
    permutation of the set $\{1,\dots N\}$,
    is the relation which
    holds if an only if
    $\textbf{e}'_{i} = \textbf{ReEnc}_{g,pk}(\textbf{e}_{\pi_M(i)},
    \textbf{r}'_{\pi_M(i)}).$
\end{itemize}

\begin{prop} 
    Algorithm 2 is a perfectly complete, 
    sound, and statistical honest verifier zero-knowledge
    4-message proof of the relationship $\mathcal{R}_{com} \vee
    (\mathcal{R}_{\pi} \underline{\land} \mathcal{R}^{shuf}_{ReEnc_{g,pk}})$.
\end{prop}

Since it is infeasible under the discrete logarithm assumption to
find a pair satisfying $\mathcal{R}_{com}$.  Thus, the
proposition computationally implies a proof of knowledge of
$\mathcal{R}_{\pi} \underline{\land} \mathcal{R}^{shuf}_{ReEnc_{g,pk}}$. 
That is for a statement $(h,h_1,...,h_N,c,g,pk,(\textbf{e}_i,...,\textbf{e}_N),(\textbf{e}'_1,...,\textbf{e}'_N))$ 
we can extract a witness $(M,\textbf{r},(\textbf{r}'_1,...,\textbf{r}'_N))$ such that 
$\mathcal{R}_{\pi}(h,h_1,...,h_N,c)(M,\textbf{r})$ and $\mathcal{R}^{shuf}_{ReEnc_{g,pk}}(g,pk,
    (\textbf{e}_1,...,\textbf{e}_N),
    (\textbf{e}'_1,...,\textbf{e}'_N))\\(\pi_M,
    (\textbf{r}'_1,...,\textbf{r}'_N))$, unless we find a discrete log.

To prove the proposition, one needs to show the correctness, the
zero-knowledge, and the soundness properties. For completeness of
the presentation, we demonstrate those properties in the
following subsections.

\subsubsection{Zero-knowledge}

    The honest-verifier zero-knowledge simulator chooses
    $\hat{c}_1,...,\hat{c}_N \in G_q$, $\hat{\textbf{s}},
    \textbf{s}', \textbf{u} \in \mathbb{Z}_q^N$, $\textbf{s}_4 \in
    \mathbb{Z}_q^w$, and $s_1,s_2,s_3, c \in \mathbb{Z}_q$ randomly
    and defines $t_1,t_2,t_3,\textbf{t}_{4,j},\hat{\textbf{t}}_i$ by
    the equations in step five. 
    
    We can observer that the statistical distance between a real
    and a simulated transcript is negligible in $q$:  

\begin{itemize}
\item 
    $\textbf{u}$ are distributed uniformly in $\mathbb{Z}_q^N$ in both.  

\item 
    $\hat{c}_1,...,\hat{c}_N$ are distributed uniformly in both
    transcripts. In the simulated one, it is easily seen by
    construction. In the real transcript
    $\hat{c}_i=g^{\hat{\textbf{r}}_i}\hat{c}_{i-1}^{\textbf{u}'_i}$,
    where $\hat{\textbf{r}}_i \in_R \mathbb{Z}_q$, which randomly
    distributes them in $G_q$ as well.

\item 
    The challenge $c$ is uniformly distributed in both

\item 
    In both transcripts, 
    $S = s_1, s_2, s_3, \textbf{s}_4, \hat{\textbf{s}}, \textbf{s}'$ 
    are distributed uniformly in their domains by their
    definitions (in the simulated
    transcript it is readily visible; in the real transcript, it
    is because $\omega$'s are distributed uniformly).

\item
    In both transcripts, the above values determine the values of
    $t_1, t_2, t_3, \textbf{t}_4, \hat{\textbf{t}}$ by
    the equations of Step 5.
\end{itemize}

\subsubsection{Correctness}

We will now show the above protocol is correct, which means that
in an honest run, the verifier accepts the proof.

We first show the shape of honest $\hat{c}_i$.

\begin{align*}
	\hat{c}_1 &= h^{\hat{\textbf{r}}_1}h_1^{\textbf{u}'_1} && \text{by definition of $\hat{c}_1$ and $\hat{c}_0$} \\
	\hat{c}_2 &= h^{\hat{\textbf{r}}_2}c_1^{\textbf{u}'_2} && \text{by definition of $\hat{c}_2$} \\
	\hat{c}_2 &= h^{\hat{\textbf{r}}_2}(h^{\hat{\textbf{r}}_1}h_1^{\textbf{u}'_1})^{\textbf{u}'_2} &&   \text{by definition of $\hat{c}_1$} \\
	\hat{c}_2 &= h^{\hat{\textbf{r}}_2 + \hat{\textbf{r}}_1 \cdot \textbf{u}'_2}h_1^{\textbf{u}'_1 \cdot \textbf{u}'_2}&&   \text{by algebraic manipulation} \\
	\hat{c}_2 &= h^{\hat{\textbf{r}}_2 + \sum_{i=1}^{2-1}\hat{\textbf{r}}_i \prod_{j=i+1}^2 \textbf{u}'_j}h_1^{\prod^2_{i=1}\textbf{u}'_i}&&   \text{by algebraic manipulation} \\
\end{align*}
Now we will continue by induction:
\begin{align*}
	\hat{c}_{\alpha} &= h^{\hat{\textbf{r}}_\alpha}\hat{c}_{\alpha-1}^{\textbf{u}'_\alpha} &&  \text{by definition of $\hat{c}_\alpha$} \\
    \hat{c}_{\alpha} &= h^{\hat{\textbf{r}}_\alpha}({h^{\hat{\textbf{r}}_{\alpha-1} + \sum_{i=1}^{\alpha-2}\hat{\textbf{r}}_i \prod_{j=i+1}^{\alpha-1} \textbf{u}'_j}h_1^{\prod^{\alpha-1}_{i=1}\textbf{u}'_i}})^{\textbf{u}'_\alpha} && \text{by definition of $\hat{c}_{\alpha-1}$ (ind.~hypothesis)}  \\
	\hat{c}_{\alpha} &= h^{\hat{\textbf{r}}_\alpha}{h^{\sum_{i=1}^{\alpha-1}\hat{\textbf{r}}_i \prod_{j=i+1}^{\alpha} \textbf{u}'_j}h_1^{\prod^{\alpha}_{i=1}\textbf{u}'_i}} && \text{by algebraic manipulation} \\
	\hat{c}_{\alpha} &= h^{\hat{\textbf{r}}_\alpha + \sum_{i=1}^{\alpha-1}\hat{\textbf{r}}_i \prod_{j=i+1}^{\alpha} \textbf{u}'_j}h_1^{\prod^{\alpha}_{i=1}\textbf{u}'_i} && \text{by algebraic manipulation} \\
\end{align*}	
Now on to the main thing. Note that in the following, we use the fact that $\textbf{c}_i$ is
a commitment to a \emph{permutation matrix} $M$ (and we will use the definition of a permutation matrix).
\begin{align*}
    t_1 &\? (\prod^N_{i=1} \textbf{c}_i / \prod^N_{i=1}h_i)^{-c}h^{s_1} &&   \text{verification definition (Step 5)} \\
	 h^{\omega_1} &\? (\prod^N_{i=1} \textbf{c}_i / \prod^N_{i=1}h_i)^{-c}h^{s_1} &&   \text{by definition of $t_1$} \\
	h^{\omega_1} &\? (\prod^N_{i=1} \textbf{c}_i / \prod^N_{i=1}h_i)^{-c}h^{ \omega_1 + c \cdot \bar{r}}&&   \text{by definition of $s_1$} \\
	 (\prod^N_{i=1} \textbf{c}_i / \prod^N_{i=1}h_i)^{c} &\? h^{c \cdot \bar{r}} &&   \text{by algebraic manipulation} \\
	  (h^{\sum^N_{i=1} \textbf{r}_i}\prod^N_{i=1}h_i/ \prod^N_{i=1}h_i)^{c} &\? h^{c \cdot \sum^N_{i=1} \textbf{r}_i} &&   \text{by definition of $\textbf{c}_i$ and $\bar{\textbf{r}}$} \\
	  h^{c \cdot \sum^N_{i=1} \textbf{r}_i} &= h^{c \cdot \sum^N_{i=1} \textbf{r}_i} &&   \text{by algebraic manipulation} \\
\end{align*}
\begin{align*}
	t_2 &\? (\hat{c}_N/h_1^{\prod^N_{i=1}\textbf{u}_i})^{-c}h^{s_2}&&   \text{verification definition} \\
	h^{\omega_2} &\? (\hat{c}_N/h_1^{\prod^N_{i=1}\textbf{u}_i})^{-c}h^{s_2}&&\text{by definition of $t_2$} \\
	h^{\omega_2} &\? (\hat{c}_N/h_1^{\prod^N_{i=1}\textbf{u}_i})^{-c}h^{ \omega_2 + c \cdot r^\diamond}&&\text{by definition of $s_2$} \\
	(\hat{c}_N/h_1^{\prod^N_{i=1}\textbf{u}_i})^{c} &\? h^{c \cdot r^\diamond}&&\text{by algebraic manipulation} \\
	(h^{\hat{\textbf{r}}_N + \sum_{i=1}^{N-1}\hat{\textbf{r}}_i \prod_{j=i+1}^{N} \textbf{u}'_j}h_1^{\prod^{N}_{i=1}\textbf{u}'_i}/h_1^{\prod^N_{i=1}\textbf{u}_i})^{c} &\? h^{c \cdot r^\diamond}&&\text{by the properties of $\hat{c}_N$} \\
	h^{c \cdot (\hat{\textbf{r}}_N + \sum_{i=1}^{N-1}\hat{\textbf{r}}_i \prod_{j=i+1}^{N} \textbf{u}'_j)} &\? h^{c \cdot r^\diamond}&&\text{by algebraic manipulation and definition of $\textbf{u}'$} \\
	h^{c \cdot (\hat{\textbf{r}}_N + \sum_{i=1}^{N-1}\hat{\textbf{r}}_i \prod_{j=i+1}^{N} \textbf{u}'_j)} &= h^{c \cdot  (\hat{\textbf{r}}_N + \sum^{N-1}_{i=1}\hat{\textbf{r}}_i\prod^{N}_{j=i+1}\textbf{u}'_{j})}&&\text{by definition $r^\diamond$} \\
\end{align*}
\begin{align*}
  t_3 &\? (\prod^N_{i=1} \textbf{c}_i^{\textbf{u}_i})^{-c} h^{s_3} \prod^N_{i=1} h_i^{\textbf{s}'_i}   &&   \text{Verification definition} \\
  	 h^{\omega_3}\prod^N_{i=1}h_i^{\bm{\omega}'_i}&\? (\prod^N_{i=1}\textbf{c}_i^{\textbf{u}_i})^{-c}h^{s_3}\prod^N_{i=1}h_i^{\textbf{s}'_i}   && \text{By definition of $t_3$} \\
  	 h^{\omega_3}\prod^N_{i=1}h_i^{\bm{\omega}'_i}&\? (\prod^N_{i=1}\textbf{c}_i^{\textbf{u}_i})^{-c}h^{\omega_3 + c \cdot \tilde{r}}\prod^N_{i=1}h_i^{ \bm{\omega}'_i + c \cdot \textbf{u}'_i}  && \text{By definition of $s_3$ and $\textbf{s}'_i$} \\
  	 (\prod^N_{i=1}\textbf{c}_i^{\textbf{u}_i})^{c} &\? h^{ c \cdot \tilde{r}}\prod^N_{i=1}h_i^{c \cdot \textbf{u}'_i} && \text{By algebraic manipulation} \\
  	 (\prod^N_{i=1}EPC{(M_i, \textbf{r}_i)^{\textbf{u}_i})}^{c} &\? EPC(c \cdot \textbf{u}', c \cdot \tilde{r})&&\text{By definition of $\textbf{c}_i$} \\ 
	(\prod^N_{i=1}EPC(M_i, \textbf{r}_i)^{\textbf{u}_i})^{c} &\? EPC(c \cdot \textbf{u}', c \cdot  \langle \textbf{r}, \textbf{u}\rangle)&&\text{By definition of $\tilde{r}$} \\
  	(\prod^N_{i=1}EPC(\textbf{u}_iM_i,\textbf{r}_i,\textbf{u}_i))^{c} &\? EPC(c \cdot \textbf{u}', c \cdot  \langle \textbf{r}, \textbf{u}\rangle)&&\text{By algebraic manipulation} \\
	 (EPC(M\textbf{u},  \langle \textbf{r}, \textbf{u}\rangle))^{c} &\? EPC(c \cdot \textbf{u}', c \cdot  \langle \textbf{r}, \textbf{u}\rangle)&&\text{By algebraic manipulation} \\
  	  (EPC( c \cdot M\textbf{u},  c \cdot \langle \textbf{r}, \textbf{u}\rangle) &\? EPC(c \cdot \textbf{u}', c \cdot  \langle \textbf{r}, \textbf{u}\rangle)&&\text{By algebraic manipulation} \\
  	  (EPC( c \cdot M\textbf{u},  c \cdot \langle \textbf{r}, \textbf{u}\rangle) &= EPC(c \cdot M\textbf{u}, c \cdot  \langle \textbf{r}, \textbf{u}\rangle)&&\text{By definition $\textbf{u}'$} \\
  \end{align*}
  \begin{align*}
	 \textbf{t}_{4} &\? \textbf{ReEnc}((\prod^N_{i=1}\textbf{e}_{i}^{\textbf{u}_i})^{-c}\prod^N_{i=1}(\textbf{e}'^{\textbf{s}'_i}_{i}),-\textbf{s}_{4})&&   \text{Verification definition} \\
	  \textbf{ReEnc}(\prod^N_{i=1} {\textbf{e}'_{i}}^{\bm{\omega}'_i}, -\bm{\omega}_{4})&\? \textbf{ReEnc}((\prod^N_{i=1}\textbf{e}_{i}^{\textbf{u}_i})^{-c}\prod^N_{i=1}(\textbf{e}'^{\textbf{s}'_i}_{i}),-\textbf{s}_{4}) && \text{By definition of $\textbf{t}_{4}$} \\
	  \prod^N_{i=1} {\textbf{e}'_{i}}^{\bm{\omega}'_i} \textbf{Enc}(1, -\bm{\omega}_{4})&\? (\prod^N_{i=1}\textbf{e}_{i}^{\textbf{u}_i})^{-c}\prod^N_{i=1}(\textbf{e}'^{\textbf{s}'_i}_{i})\textbf{Enc}(1,-\textbf{s}_{4}) && \text{By definition of \textbf{ReEnc}} \\
	   \prod^N_{i=1} {\textbf{e}'_{i}}^{\bm{\omega}'_i} \textbf{Enc}(1, -\bm{\omega}_{4})&\? (\prod^N_{i=1}\textbf{e}_{i}^{\textbf{u}_i})^{-c}\prod^N_{i=1}(\textbf{e}'^{\textbf{s}'_i}_{i})\textbf{Enc}(1,-\bm{\omega}_{4} - c \cdot \textbf{r}^\star) && \text{By definition of $\textbf{s}_{4}$} \\
	   \prod^N_{i=1} {\textbf{e}'_{i}}^{\bm{\omega}'_i}&\? (\prod^N_{i=1}\textbf{e}_{i}^{\textbf{u}_i})^{-c}\prod^N_{i=1}(\textbf{e}'^{\textbf{s}'_i}_{i})\textbf{Enc}(1,- c \cdot \textbf{r}^\star) && \text{By algebraic manipulation} \\
	   \prod^N_{i=1} {\textbf{e}'_{i}}^{\bm{\omega}'_i}&\? (\prod^N_{i=1}\textbf{e}_{i}^{\textbf{u}_i})^{-c}\prod^N_{i=1}(\textbf{e}'^{\bm{\omega}'_i + c \cdot \textbf{u}'_i }_{i,j})\textbf{Enc}(1,- c \cdot \textbf{r}^\star) && \text{By definition of $\textbf{s}'_i$} \\
	   (\prod^N_{i=1}\textbf{e}_{i}^{\textbf{u}_i})^{c} &\? (\prod^N_{i=1}{\textbf{e}'_{i}}^{\textbf{u}'_i})^{c}\textbf{Enc}(1,- c \cdot \textbf{r}^\star) && \text{By algebraic manipulation} \\
	   (\prod^N_{i=1}\textbf{e}_{i}^{\textbf{u}_i})^{c} &\? (\prod^N_{i=1}(\textbf{e}_{i}\textbf{Enc}(1,R_{i}))^{\textbf{u}_i} )^{c}\textbf{Enc}(1,- c \cdot \textbf{r}^\star) && \text{By definition of $\textbf{e}'$ and $\textbf{u}'$} \\
	   \textbf{Enc}(1, c \cdot \textbf{r}^\star) &\? (\prod^N_{i=1}\textbf{Enc}_{pk}(1,R_{i})^{\textbf{u}_i} )^{c} && \text{By algebraic manipulation} \\
	   \textbf{Enc}(1,  c \cdot \textbf{r}^\star) &\? \textbf{Enc}(1, c \cdot R\textbf{u}) && \text{By algebraic manipulation} \\
	   \textbf{Enc}(1,  c \cdot \textbf{r}^\star) &= \textbf{Enc}(1, c \cdot \textbf{r}^\star) && \text{By definition of $\textbf{r}^\star$} \\
\end{align*}
  \begin{align*} 
	  \hat{\textbf{t}}_i &\? \hat{c}_i^{-c}h^{\hat{\textbf{s}}_i}\hat{c}_{i-1}^{\textbf{s}'_i}&&   \text{Verification definition} \\
	   h^{\bm{\hat{\omega}}_i}\hat{c}_{i-1}^{\bm{\omega}'_i} &\? \hat{c}_i^{-c}h^{\hat{\textbf{s}}_i}\hat{c}_{i-1}^{\textbf{s}'_i} &&   \text{By definition of $\hat{\textbf{t}}_i$} \\
	   h^{\bm{\hat{\omega}}_i}\hat{c}_{i-1}^{\bm{\omega}'_i} &\? \hat{c}_i^{-c}h^{ \bm{\hat{\omega}}_i + c \cdot \hat{\textbf{r}}_i}\hat{c}_{i-1}^{\bm{\omega}'_i + c \cdot \textbf{u}'_i} &&   \text{By definition of $\hat{\textbf{s}}_i$ and $\textbf{s}'_i$} \\
	   \hat{c}_i^{c} &\? h^{c \cdot \hat{\textbf{r}}_i}\hat{c}_{i-1}^{c \cdot \textbf{u}'_i} && \text{By algebraic manipulation} \\
	   (h^{\hat{\textbf{r}}_i}\hat{c}_{i-1}^{\textbf{u}'_i})^{c} &\? h^{c \cdot \hat{\textbf{r}}_i}\hat{c}_{i-1}^{c \cdot \textbf{u}'_i} && \text{By definition of $\hat{c}_i$} \\
	    h^{c \cdot \hat{\textbf{r}}_i}\hat{c}_{i-1}^{c \cdot \textbf{u}'_i}  &=  h^{c \cdot \hat{\textbf{r}}_i}\hat{c}_{i-1}^{c \cdot \textbf{u}'_i}   && \text{By algebraic manipulation} \\
\end{align*}

\subsubsection{Soundness}

We follow the structure of the original proof, as presented in
\cite{TereliusWikstroem-AFRICACRYPT-2010}, and present the extractor in two parts. First, we show
that, for two different transcripts with the same $\textbf{u}$ but different
$c$, we can extract witness for certain sub-statements.  In the extended
extractor we show that, given witnesses for these sub-statements which hold for
$n$ different $\textbf{u}$, we can extract witness to the main statements.

\paragraph{Basic extractor.}
Given two accepting transcripts
\begin{align*}
    &(\textbf{u},\hat{\textbf{c}},t_1,t_2,t_3,\textbf{t}_4,\hat{\textbf{t}},c,s_1,s_2,s_3,\textbf{s}_4,\hat{\textbf{s}},\textbf{s}') \\
    &(\textbf{u},\hat{\textbf{c}},t_1,t_2,t_3,\textbf{t}_4,\hat{\textbf{t}},c^*,s_1^*,s_2^*,s_3^*,\textbf{s}_4^*,\hat{\textbf{s}}^*,\textbf{s}'^*)
\end{align*}
with $c \neq c^*$, the \emph{basic extractor} computes 
\begin{align*}
    \bar{r} &= (s_1-s_1^*)/(c - c^*), 
    &r^\diamond &= (s_2-s_2^*)/(c - c^*), \\
    \tilde{r} &= (s_3-s_3^*)/(c - c^*),
    &\textbf{r}^\star &= (\textbf{s}_{4}-\textbf{s}_{4}^*)/(c - c^*), \\
    \hat{\textbf{r}} &= (\hat{\textbf{s}} - \hat{\textbf{s}}^*)/(c - c^*),
    &\textbf{u}' &= (\textbf{s}' - \textbf{s}'^*)/(c - c^*).
\end{align*}
Note that we reuse symbols from the Algorithm~\ref{algo:shuffle}. While they
denote analogous entities, they are not necessarily identical (if the transcripts
have not been obtained in the honest way).

We will prove that
\[
	\prod^{N}_{j=1} \textbf{c}_j = EPC(\bm{1},\bar{r}) \hspace{0.5cm} 
	 \prod^{N}_{j=1} \textbf{c}_j^{\textbf{u}_j} = EPC(\bm{u}', \tilde{r})  \hspace{0.5cm} 
	\prod^{N}_{i=1} {\textbf{e}'_{i}}^{\textbf{u}'_i} = \textbf{Enc}_{pk}(1,\textbf{r}^\star) \cdot \prod^{N}_{i=1} \textbf{e}_{i}^{\textbf{u}_i}
\]
\[
	\hat{c}_i = PC_{h,\hat{c}_{i-1}}(\textbf{u}'_i,\hat{\textbf{r}}_i) \hspace{0.5cm}  \hat{c}_N = PC_{h,h_1}(\prod_{i=1}^N \textbf{u}_i, r^\diamond) 
\]
The proof consists of simple algebraic transformations:
\begin{align*}
    \left(\frac{(\prod^{N}_{j=1} \textbf{c}_j)^c t_1 }{(\prod^{N}_{j=1} \textbf{c}_j)^{c^*} t_1}\right)^{\frac{1}{c-c^*}} 
        =\prod^{N}_{j=1} \textbf{c}_j && \text{Tautology} \\
    \left(\frac{h^{s_1}/ (\prod^N_{i=1}h_i)^{-c} }{h^{s_1^*}/ (\prod^N_{i=1}h_i)^{-c^*}}\right)^{\frac{1}{c-c^*}}
        =\prod^{N}_{j=1} \textbf{c}_j && \text{By the verification definition} \\
	h^{\frac{s_1-s^*_1}{c-c^*}}\prod^N_{i=1}h_i=\prod^{N}_{j=1} \textbf{c}_j && \text{By algebraic manipulation} \\
	EPC(\textbf{1},\frac{s_1-s^*_1}{c-c^*}) = \prod^{N}_{j=1} \textbf{c}_j && \text{By definition of EPC} \\
	EPC(\textbf{1},\bar{r}) = \prod^{N}_{j=1} \textbf{c}_j && \text{By definition of $\bar{r}$}
\end{align*}
\begin{align*}
    \left(\frac{(\prod^{N}_{j=1} \textbf{c}_j^{\textbf{u}_j})^ct_3}{(\prod^{N}_{j=1} \textbf{c}_j^{\textbf{u}_j})^{c^*}t_3} \right)^{\frac{1}{c-c^*}} 
        = \prod^{N}_{j=1} \textbf{c}_j^{\textbf{u}_j} && \text{Tautology} \\
    \left(\frac{h^{s_3}\prod^N_{i=1}h_i^{\textbf{s}'_i}}{h^{s_3^*}\prod^N_{i=1}h_i^{\textbf{s}'^*_i}}\right)^{\frac{1}{c-c^*}} = \prod^{N}_{j=1} \textbf{c}_j^{\textbf{u}_j}  && \text{By verification definition} \\
	h^{{\frac{s_3-s_3^*}{c-c^*}} }\prod^N_{i=1}h_i^{\frac{\textbf{s}'_i-\textbf{s}'^*_i}{c-c^*}}= \prod^{N}_{j=1} \textbf{c}_j^{\textbf{u}_j} && \text{By algebraic manipulation} \\
	EPC(\frac{\textbf{s}'-\textbf{s}'^*}{c-c^*},\frac{s_3-s_3^*}{c-c^*}) = \prod^{N}_{j=1} \textbf{c}_j^{\textbf{u}_j} && \text{By definition of EPC} \\
	EPC(\bm{u}', \tilde{r}) = \prod^{N}_{j=1} \textbf{c}_j^{\textbf{u}_j} && \text{By definition of $\bm{u}'$ and $\tilde{r}$}
\end{align*}
\begin{align*}
    \left(\frac{(\prod^{N}_{i=1} (\textbf{e}_{i})^{\textbf{u}_i})^c\textbf{t}_{4}}{(\prod^{N}_{i=1} {(\textbf{e}_{i})}^{\textbf{u}_i})^{c^*}\textbf{t}_{4}} \right) ^{\frac{1}{c-c^*}} 
        &= \prod^{N}_{i=1} {\textbf{e}_{i}}^{\textbf{u}_i}  && \text{Tautology}  \\
    \left(\frac{\prod^N_{i=1}(\textbf{e}'_{i})^{\textbf{s}'_i}\textbf{Enc}(1,-\textbf{s}_{4})}{\prod^N_{i=1}(\textbf{e}'_{i})^{\textbf{s}'^*_i}\textbf{Enc}(1,-\textbf{s}_{4}^*)}\right)^{\frac{1}{c-c^*}}  &= \prod^{N}_{i=1} {\textbf{e}_{i}}^{\textbf{u}_i} && \text{By verification definition} \\
	 \prod^N_{i=1}\textbf{e}'^{\frac{\textbf{s}'_i-\textbf{s}'^*_i}{c-c^*}}_{i}\textbf{Enc}(1,\frac{\textbf{s}_{4}^*-\textbf{s}_{4}}{c-c^*}) &= \prod^{N}_{i=1} {\textbf{e}_{i}}^{\textbf{u}_i} && \text{By algebraic manipulation} \\
	  \prod^N_{i=1}\textbf{e}'^{\frac{\textbf{s}'_i-\textbf{s}'^*_i}{c-c^*}}_{i} &= \textbf{Enc}(1,\frac{\textbf{s}_{4}-\textbf{s}_{4}^*}{c-c^*})\prod^{N}_{i=1} {\textbf{e}_{i}}^{\textbf{u}_i} && \text{By algebraic manipulation} \\
	   \prod^{N}_{i=1} {\textbf{e}'_{i}}^{\textbf{u}'_i} &= \textbf{Enc}_{pk}(1,\textbf{r}^\star)  \prod^{N}_{i=1} {\textbf{e}_{i}}^{\textbf{u}_i} && \text{By definition of $r'_j$ and $\textbf{u}'_i$} \\
\end{align*}
Now, for each $i\in\{1,\dots,N\}$
\begin{align*}
    \left(\frac{\hat{c}_i^c\hat{\textbf{t}}_i}{\hat{c}_i^{c^*}\hat{\textbf{t}}_i}\right)^{\frac{1}{c-c^*}} = \hat{c}_i && \text{Tautology} \\
    \left(\frac{h^{\hat{\textbf{s}}_i}\hat{c}_{i-1}^{\textbf{s}'_i}}{h^{\hat{\textbf{s}}_i^*}\hat{c}_{i-1}^{\textbf{s}'^*_i}}\right)^{\frac{1}{c-c^*}} = \hat{c}_i && \text{By verification definition} \\
	h^{\frac{\hat{\textbf{s}}_i-\hat{\textbf{s}}_i^*}{c-c^*}}\hat{c}_{i-1}^{\frac{\textbf{s}'_i-\textbf{s}'^*_i}{c-c^*}} = \hat{c}_i && \text{By algebraic manipulations} \\
	PC_{h,\hat{c}_{i-1}}(\frac{\textbf{s}'_i-\textbf{s}'^*_i}{c-c^*},\frac{\hat{\textbf{s}}_i-\hat{\textbf{s}}_i^*}{c-c^*})  =  \hat{c}_i && \text{By algebraic manipulations} \\
	PC_{h,\hat{c}_{i-1}}(\textbf{u}'_i,\hat{\textbf{r}}_i) = \hat{c}_i && \text{By definition of $\textbf{u}'_i$ and $\hat{\textbf{r}}_i$}  \\
\end{align*}
\begin{align*}
    \left(\frac{\hat{c}_N^ct_2}{\hat{c}_N^{c^*}t_2}\right)^{\frac{1}{c-c^*}} = \hat{c}_N && \text{Tautology} \\
    \left(\frac{(h_1^{\prod^N_{i=1}\textbf{u}_i})^{c}h^{s_2}}{(h_1^{\prod^N_{i=1}\textbf{u}_i})^{c^*}h^{s_2^*}}^{\frac{1}{c-c^*}}\right) = \hat{c}_N && \text{By verification definition} \\
	h^{\frac{s_2-s_2^*}{c-c^*}}h_1^{\prod^N_{i=1}\textbf{u}_i} = \hat{c}_N && \text{By algebraic manipulation} \\
	PC(\prod_{i=1}^N \textbf{u}_i,\frac{s_2-s_2^*}{c-c^*}) = \hat{c}_N && \text{By algebraic manipulation}  \\
	PC_{h,h_1}(\prod_{i=1}^N \textbf{u}_i, r^\diamond)  = \hat{c}_N && \text{By definition of $r^\diamond$}  \\
\end{align*}

\paragraph{Extended Extractor}\label{subsec:exex}

We now sketch the extended extractor which, for a given statement (see the
common input in Algorithm~\ref{algo:shuffle}), for $n$ different witnesses
extracted by the basic extractor, produces the witnesses to the main statement.
Let the collective output of the basic extractors be denoted as
$\bar{\textbf{r}}, \textbf{r}^\diamond, \tilde{\textbf{r}}  \in \ZZ_q^n$,
$R^\star \in \ZZ_q^{W\times N}$, and $\hat{R}, U' \in
\ZZ_q^{N \times N}$ extracted from the primary challenges $U \in
\ZZ_q^{N\times N}$.  We denote by $U_i$ the $i$th column of
$U$ which is the challenge vector from the $i$th run of the basic
extractor, and by $U_{j,i}$ the $j$ element of the challenge vector
from the $i$th run of the basic extractor. 

First note with overwhelming probability the set of $U_i$s is linearly
independent, concretely the probability is bounded by $\frac{q-2}{q}$.  From
linear independence, it follows that their exists $A \in \ZZ_q^{N \times N}$ such
that  $UA_{l}$ is the $l$th standard unit
vector in $\mathbb{Z}_q$ which we will denote by $\mathbb{I}_l$.  $A$ is the inverse of 
$U$.  Clearly,

\begin{align*}
    \textbf{c}_l &= \prod_{i=1}^N (\textbf{c}^{UA_l})_i && \text{since $UA_{l}$ is $\mathbb{I}_l$} \\
    \textbf{c}_l &= \prod_{i=1}^N \textbf{c}_i^{\sum_{j=1}^N U_{i,j} A_{j,l}} && \text{by definition of $UA_l$} \\
    \textbf{c}_l &= \prod_{i=1}^N\left(\prod_{j=1}^N \textbf{c}^{U_{i,j}A_{j,l}}\right)_i && \text{by algebraic manipulation}  \\ 
    \textbf{c}_l &= \prod_{j=1}^N \left(\left(\prod_{i=1}^N \textbf{c}_i^{U_{i,j}}\right)^{A_{j,l}}\right)&& \text{by algebraic manipulation} \\
    \textbf{c}_l &= \prod_{j=1}^N EPC(U'_j,\tilde{\textbf{r}}_j)^{A_{j,l}} && \text{by some algebraic manipulation and $\prod_{i=1}^N\textbf{c}_{i}^{U_{i,j}} = EPC(U'_j, \tilde{\textbf{r}}_j)$} \\
    \textbf{c}_l &= \prod^N_{j=1}EPC(U'_jA_{j,l},  \tilde{\textbf{r}}_j A_{j,l}) && \text{by algebraic manipulation} \\
    \textbf{c}_l &= EPC(\sum_{j=1}^N U'_jA_{j,l},  \langle \tilde{\textbf{r}}, A_{l} \rangle) && \text{by algebraic manipulation} \\
    \textbf{c}_l &= EPC(U'A_{l},  \langle \tilde{\textbf{r}}, A_{l} \rangle) && \text{by algebraic manipulation}
\end{align*}
Therefore, we can open $\textbf{c}$ to the matrix $M$, where the $l$th column of $M$ is
$U'A_{l}$, with randomness
$\langle \tilde{\textbf{r}}, A_{l} \rangle$.  In other words we open $\textbf{c} = U'A$ using randomness $\tilde{\textbf{r}}A$.

We expect $M$ to be a permutation matrix, but if it is not, then one can
find a witness to $\mathcal{R}_{com}$ (which, as has been mentioned, can only
happen with negligible probability, under our security assumptions). 
We extract in two different ways depending on whether $M\textbf{1} \neq \textbf{1}$.

\paragraph{Option one}
If $M\textbf{1} \neq \textbf{1}$, then let $\textbf{u}'' = M\textbf{1}$ and note that 
\begin{align*}
	\textbf{u}'' \neq \textbf{1} \text{ and } EPC(\textbf{1}, \bar{\textbf{r}}_j) = \prod_{i=1}^N \textbf{c}_i = \prod_{i=1}^N\textbf{c}_i^{\textbf{1}_i} = EPC(\textbf{u}'', \tilde{\textbf{r}}A)
\end{align*}
in which case we found a witness breaking the commitment scheme.

\paragraph{Option two}
If $M\textbf{1} = \textbf{1}$, then recall Theorem 1 from ``Proofs of Restricted Shuffles'', which states that $M$ is a permutation matrix if and only if
$M\textbf{1} = \textbf{1}$ and $\prod^N_{i=1} \langle \textbf{m}_i, \textbf{x} \rangle - \prod^N_{i=1} \textbf{x}_i = 0$.
Since $M\textbf{1} = \textbf{1}$ and $M$ is not a permutation matrix, then $\prod^N_{i=1} \langle \textbf{m}_i, \textbf{x} \rangle - \prod^N_{i=1} \textbf{x}_i \neq 0$.  
The Schwartz–Zippel says that if you sample, a non-zero polynomial, at a random point the chance that it equals zero is negligible in the order of the underlying field; hence, with overwhelming probability there exists $j\in\{1,\dots, N\}$ such $\prod^N_{i=1}
\langle \textbf{m}_i, U_j \rangle - \prod^N_{i=1} U_{i,j} \neq 0$.
Since this is true with overwhelming probability, we require it to be true and
rewind if this is not the case.  (Strictly speaking we should take $N+1$ extractions from the basic extractor, if we recover a different $M$ we win, if we get the same $M$ then $U_{l+1}$ is actually independent of $M$ and the lemma can be applied.)  \newline \newline

Let $\textbf{u}'' = M U_j$ and note that
\begin{align*}
	\textbf{u}'' \neq U'_j && \text{Which must be true since $\prod^N_{i=1} U'_{i,j} = \prod^N_{i=1} U_{i,j} \neq \prod^N_{i=1} \textbf{u}''_i$} \\
\end{align*}
$\prod^N_{i=1} U'_j = \prod^N_{i=1} U_j$ follows from the base statements and $\prod^N_{i=1} U_j \neq \prod^N_{i=1} \textbf{u}''$ by definition of $\textbf{u}''$ and $\prod^N_{i=1}
\langle \textbf{m}_i, U_j \rangle - \prod^N_{i=1} U_{i,j} \neq 0$.
\begin{align*}
	EPC(U'_j, \tilde{\textbf{r}}_j) = \prod_{i=1}^N \textbf{c}_i^{U_{i,j}} = EPC(\textbf{u}'',  \langle  \tilde{\textbf{r}A}, U_j \rangle) && \text{} \\
\end{align*}
This completes the proof that $M$ is a permutation matrix or we have found a witness to $\mathcal{R}_{com}$.

\paragraph{The correctness of $U'$}

We now show that $U'_l = MU_l$ for all $l  \in [1,N]$ or we can find a witnesses to $\mathcal{R}_{com}$.
  Let $\textbf{u}'' = MU_l$ and by assumption $\textbf{u}'' \neq U'_l$.  
\begin{align*}
	EPC(U'_l, \tilde{\textbf{r}}_l) = \prod_{i=1}^N \textbf{c}_i^{U_{i,l}} = EPC(\textbf{u}'',  \langle \tilde{\textbf{r}}A, U_l \rangle) && \text{} \\
\end{align*}
\paragraph{Extracting the randomness} We having shown that if $M$ is not a permutation matrix we can extract a witness to $\mathcal{R}_{com}$.  We now show that we can extract $R \in \ZZ_q^{w \times N}$
    such that 
    $\textbf{e}'_{i} = \textbf{ReEnc}_{pk}(\textbf{e}_{\pi(i)},R_{\pi(i)})$.  
\begin{align}
	\textbf{e}_l &= \prod_{i=1}^N (\textbf{e}^{UA_l})_i && \text{since $UA_{l}$ is $\mathbb{I}_l$} \\
        \textbf{e}_l &= \prod_{i=1}^N \textbf{e}_i^{\sum_{j=1}^N U_{i,j} A_{j,l}} && \text{by definition of $UA_l$} \\
        \textbf{e}_l &= \prod_{i=1}^N\left( \prod_{j=1}^N \textbf{e}^{U_{i,j}A_{j,l}}\right)_i && \text{by algebraic manipulation}  \\ 
	\textbf{e}_l &= \prod_{j=1}^N\left( \prod_{i=1}^N \textbf{e}^{U_{i,j}}\right)^{A_{j,l}} && \text{by algebraic manipulation} \\
	\textbf{e}_l &= \prod_{j=1}^N\left(\prod_{i=1}^N {\textbf{e}'_{i}}^{U'_{i,j}}\textbf{Enc}_{pk}(1,-R^*_j)\right)^{A_{j,l}} &&
	 \text{since $\prod_{i=1}^N \textbf{e}_i^{U_{i,j}} = \prod_{i=1}^N {\textbf{e}'_i}^{U'_{i,j}}\textbf{Enc}_{pk}(1,-R^\star_j)$} \\
	\textbf{e}_l &= \prod_{j=1}^N \left(\prod_{i=1}^N {\textbf{e}'_{i}}^{U'_{i,j}A_{j,l}}\textbf{Enc}_{pk}(1,-R^*_jA_{j,l})\right) && \text{by algebraic manipulation} \\
	\textbf{e}_l  &= \prod_{i=1}^N {\textbf{e}'_{i}}^{\sum_{j=1}^N U'_{i,j}A_{j,l}\textbf{Enc}_{pk}}(1,-\langle R^*, A_{l} \rangle) && \text{by algebraic manipulation} \\
	\textbf{e}_l  &= \prod_{i=1}^N(\textbf{e}'^{U'A_l})_i\textbf{Enc}_{pk}(1,-\langle R^*, A_{l} \rangle) && \text{by algebraic manipulation} \\
	\textbf{e}_l  &= \prod_{i=1}^N(\textbf{e}'^{MUA_l})_i\textbf{Enc}_{pk}(1,-\langle R^*, A_{l} \rangle) && \text{since U' = MU}  \\
	\textbf{e}_l  &= \prod_{i=1}^N(\textbf{e}'^{M\mathbb{I}_l})_i\textbf{Enc}_{pk}(1,-\langle R^*, A_{l} \rangle) && \text{since $UA_l = \mathbb{I}_l$}  \\
	\textbf{e}_l  &= \prod_{i=1}^N(\textbf{e}'^{M_l})_i\textbf{Enc}_{pk}(1,-\langle R^*, A_{l} \rangle) && \text{since $M\mathbb{I}_l = M_l$}  \\
	\textbf{e}_l  &= \textbf{e}'_{\pi_M^{-1}(l)}\textbf{Enc}_{pk}(1,-\langle R^*, A_{l} \rangle) && \text{by definition of $\pi_M$}  \\
\end{align}
We have now shown that $\textbf{ReEnc}_{pk}(\textbf{e}_l,\langle R^*_l, A_{l} \rangle) = \textbf{e}'_{\pi^{-1}(l)}$; hence, $R_l = \langle R^*_l, A_{l} \rangle$  which concludes the proof.

\end{document}